\documentclass[journal=nalefd,manuscript=letter]{achemso}

\usepackage[T1]{fontenc}
\usepackage{lmodern}
\usepackage{amstext}
\usepackage{amsmath}
\usepackage{siunitx}
\usepackage{graphicx}
\usepackage{subfigure}
\usepackage{xcolor}
\usepackage{wasysym}
\usepackage[export]{adjustbox}
\usepackage{hyperref}
\usepackage{amsmath, amssymb}
\usepackage{color,soul}
\usepackage{tabularx}

\usepackage{caption}
\usepackage{setspace}

\usepackage{romanbar}
\usepackage{textcomp, amssymb}
\usepackage{comment} 
\usepackage{booktabs}
\usepackage{lipsum}
\usepackage{enumitem}
\usepackage{wrapfig}

\author{Mahmoud Al Humaidi }
\affiliation[University of Siegen]{University of Siegen, Solid State Physics, Emmy-Noether Campus, Walter-Flex Stra\ss{}e 3, D-57068 Siegen, Germany}
\alsoaffiliation[KIT-LAS]{Karlsruhe Institute of Technology, Laboratory for Applications of Synchrotron Radiation, Kaiserstra\ss{}e 12, D-76131 Karlsruhe, Germany}
\email{mahmoud.humaidi@kit.edu}

\author{Julian Jakob}
\affiliation[KIT-IPS]{Karlsruhe Institute of Technology, Institute for Photon Science and Synchrotron Radiation, Hermann-von-Helmholtz-Platz 1, D-76344 Eggenstein-Leopoldshafen, Germany}
\alsoaffiliation[KIT-LAS]{Karlsruhe Institute of Technology, Laboratory for Applications of Synchrotron Radiation, Kaiserstra\ss{}e 12, D-76131 Karlsruhe, Germany}

\author{Ali Al Hassan}
\alsoaffiliation[KIT-IPS]{Karlsruhe Institute of Technology, Institute for Photon Science and Synchrotron Radiation, Hermann-von-Helmholtz-Platz 1, D-76344 Eggenstein-Leopoldshafen, Germany}

\author{Arman Davtyan}
\affiliation[University of Siegen]{University of Siegen, Solid State Physics, Emmy-Noether Campus, Walter-Flex Stra\ss{}e 3, D-57068 Siegen, Germany}

\author{Philipp Schroth}
\affiliation[KIT-IPS]{Karlsruhe Institute of Technology, Institute for Photon Science and Synchrotron Radiation, Hermann-von-Helmholtz-Platz 1, D-76344 Eggenstein-Leopoldshafen, Germany}
\alsoaffiliation[KIT-LAS]{Karlsruhe Institute of Technology, Laboratory for Applications of Synchrotron Radiation, Kaiserstra\ss{}e 12, D-76131 Karlsruhe, Germany}

\author{Ludwig Feigl}
\affiliation[KIT-IPS]{Karlsruhe Institute of Technology, Institute for Photon Science and Synchrotron Radiation, Hermann-von-Helmholtz-Platz 1, D-76344 Eggenstein-Leopoldshafen, Germany}

\author{Jes\'{u}s Herranz}
\affiliation{Paul-Drude-Institut f\"{u}r Festk\"{o}rperelektronik, Leibniz-Institut im Forschungsverbund Berlin e.V., Hausvogteiplatz 5-7, 10117 Berlin, Germany}

\author{Dmitri Novikov}
\affiliation{Deutsches Elektronen-Synchrotron, PETRA III, Hamburg D-22607, Germany}

\author{Lutz Geelhaar}
\affiliation{Paul-Drude-Institut f\"{u}r Festk\"{o}rperelektronik, Leibniz-Institut im Forschungsverbund Berlin e.V., Hausvogteiplatz 5-7, 10117 Berlin, Germany}

\author{Tilo Baumbach}
\affiliation[KIT-IPS]{Karlsruhe Institute of Technology, Institute for Photon Science and Synchrotron Radiation, Hermann-von-Helmholtz-Platz 1, D-76344 Eggenstein-Leopoldshafen, Germany}
\alsoaffiliation[KIT-LAS]{Karlsruhe Institute of Technology, Laboratory for Applications of Synchrotron Radiation, Kaiserstra\ss{}e 12, D-76131 Karlsruhe, Germany}

\author{Ullrich Pietsch}
\affiliation[University of Siegen]{University of Siegen, Solid State Physics, Emmy-Noether Campus, Walter-Flex Stra\ss{}e 3, D-57068 Siegen, Germany}

\title{Exploiting of flux shadowing effect on In$_{x}$Ga$_{1-x}$As asymmetric shell growth for strain and bending engineering in GaAs - In$_{x}$Ga$_{1-x}$As core - shell NW arrays}

\begin{document}

\date{\today}

\newpage

\begin{abstract}
Here we report on non-uniform shell growth of In$_x$Ga$_{1-x}$As onto GaAs nanowire (NW) core by molecular beam epitaxy (MBE). The growth was realized on pre-patterned silicon substrates with pitch size ($p$) ranging from 0.1 µm to 10 µm. Considering the preferable bending direction with respect to the MBE cells as well as the layout of the substrate pattern, we are able to modify the strain distribution along the NW growth axis and the subsequent bending profile. For NW arrays with high number density, the obtained bending profile of the NWs is composed of straight (barely-strained) and bent (strained) segments with different lengths which depend on the pitch size. A precise control of the bent and straight NW segment length provides a recipe to design NW based devices with length selective strain distribution. 
\end{abstract}

\maketitle

\newpage
\subsubsection{Introduction}
Due to their large surface-to-volume ratio, NWs have size-dependent mechanical properties allowing for strain engineering\cite{NasrEsfahani2019}. Since piezoelectric fields in strained NWs can be utilized for efficient carrier sweeping toward device electrodes\cite{Boxberg2010,Fu}, strain engineering can be used to tune the electronic band gap and to tailor the performance of NW-based devices\cite{Lee2005,Lim2021,Bartmann2021}. Strained heteroepitaxial core-shell NWs formed from lattice-mismatched materials have potential applications in NW-based devices such as light-emitting diodes\cite{Tatebayashi2010,Dalacu2019}, solar cells\cite{Tang2011,Goto2009,Han2015,Moratis2016} and electronic devices \cite{Jiang2007,Ieong}. 
As a consequence of the elastic release of interface strain, thin NWs bend if a heteroepitaxial shell is grown inhomogeneously on their perimeter\cite{Gagliano2018,Hilse2009}. In case of growth of a shell with higher lattice parameter compared to the core, the strain gradient across the NW diameter vary from tensile strain at the shell surface towards compressive strain at the opposite side\cite{Lewis2018,Greenberg2019}. Due to the impact of deformation potentials on the band structure the strain gradient across the bent NW induces a drift of the charge carriers toward the regions with tensile strain which allows to tune the electronic properties of the NWs for future optoelectronic devices. Moreover, bent NWs can be used for NW networks or interconnects for electronics with novel designs\cite{Jiang2007}. Lewis et {$al$}. \cite{Lewis2018} reported on NW bending induced by growing a non-uniform Al$_{0.5}$In$_{0.5}$As shell ($\approx$ 3.6\% lattice mismatch) onto GaAs NW cores. The non-uniformity of the shell growth in this case was achieved by sequential NWs rotation and shell deposition and resulted in shell growth only on defined NW sides. Recently, we have reported on non-uniform shell growth by MBE deposition without substrate rotation where the shell was grown mainly on NW side walls defined by a certain flux direction of the growth materials \cite{Alhumaidi2021}. Controlling core diameter, $d$, shell thickness, $t$, and alloy concentration, $x$, in the shell, it was possible to manipulate the curvature of the NW. However, the NW growth reported in\cite{Lewis2018} was performed on silicon substrate with low number density resulting in homogeneous bending of all NWs along the entire length. In this work, we report on GaAs NW growth on pre-patterned Si(111) substrates followed by a lattice mismatched In$_{x}$Ga$_{1-x}$As shell growth without substrate rotation. To control the axial distribution of the shell material along the NW, we benefit from the well-defined pattern of the NWs and their volume with respect to the MBE geometry. We exploit the shadowing of these fluxes by neighboring NWs and consequently a varying strain distribution along to obtain different bending profile of the NWs. 
Our findings are observed by \textit{in-situ} X-ray diffraction (XRD) measurements performed at NW arrays with different pitch size (i.e. different number density of the NWs) during shell growth.

\section{The fundamental idea and experimental methods}
In this study we used two samples of NW arrays where the separation between neighboring NWs (pitch $p$) of the patterned arrays differs for each array, ranging from $p= 0.1$ µm to $10$ µm. The shell growth parameters for sample-1 were chosen in a way to have higher NW bending comparing to sample-2 to demonstrate the effect of flux shadowing by scanning electron microscopy (SEM). The In content for sample-1 was $x=0.3$ and the shell growth time was 30 minutes. During shell growth, sample-1 was azimuthally aligned in a way that the flux of the shell materials was shadowed by the nearest neighboring NWs (i.e. the distance between the NWs on the bending direction is $p$). On the other hand, sample-2 was grown to monitor the evolution of the strain and NW bending at different NW arrays with different $p$ during shell growth by {$in-situ$} XRD experiment. Therefore, sample-2 was grown with lower In content ($x=0.15$) and a shell growth time of 20 minutes, resulting in a lower bending compared to sample-1. The shadowing in sample-2 is done by the next neighboring NWs (i.e. the distance between the NWs along the bending direction is $\sqrt{3}\:p$ where the arrays pattern has a hexagonal grid). The shell growth for sample-2 was realized in several steps as we deposited the shell material for certain time intervals, followed by an interruption of the growth in which the XRD experiment is performed. %

\begin{figure}[ht!]
	\centering
	\includegraphics[width=13cm]{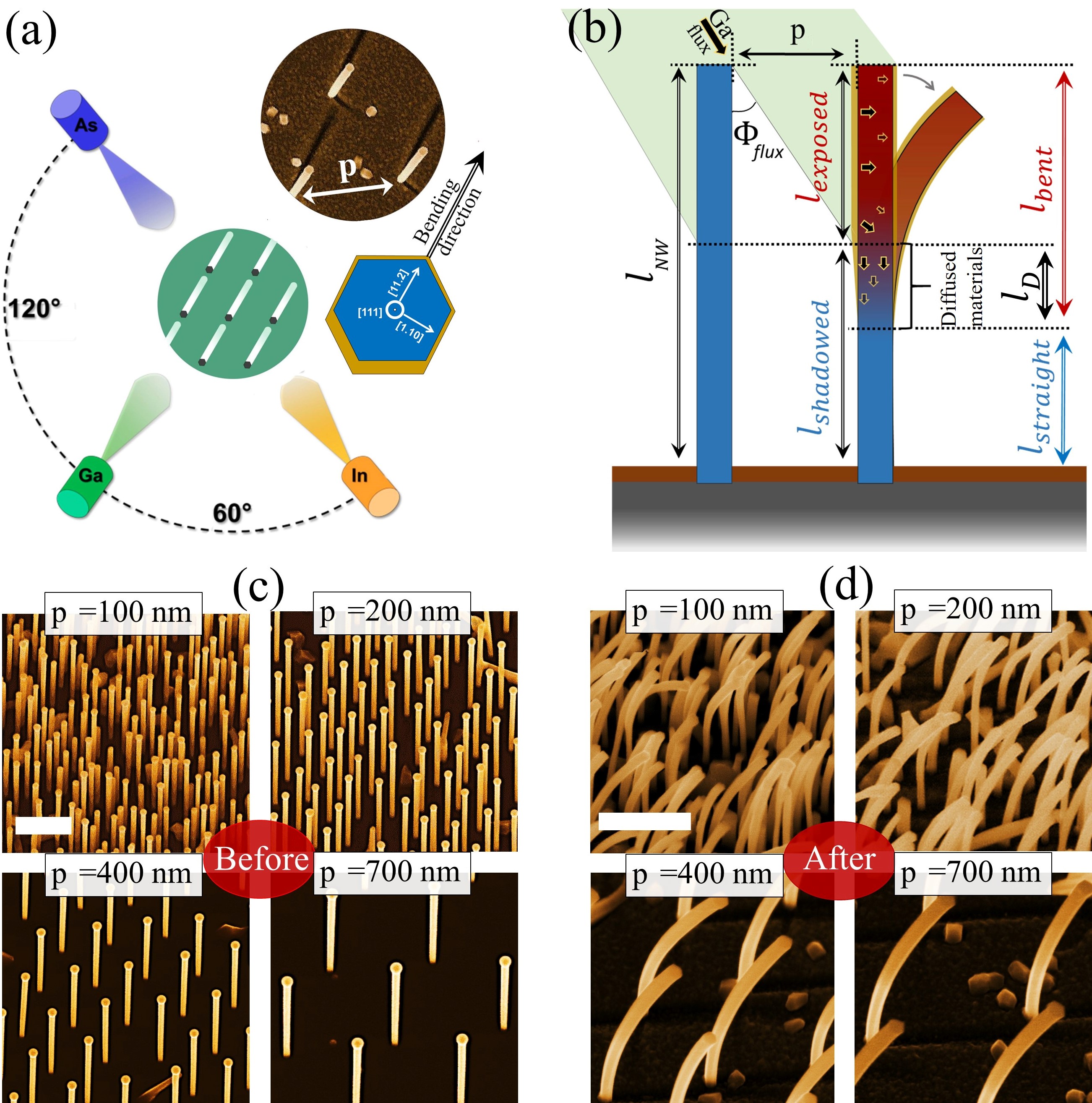}
	\caption{ (a) Illustration of the azimuthal arrangement of the MBE cells and the pattern of the substrate. (b) Illustration of the materials flux shadowing by neighbouring NWs. (c) 30° tilt view SEM images of reference GaAs NW arrays with different pitch size $p$ before shell growth. (d) 30° tilt view SEM images of bent GaAs - In$_{0.3}$Ga$_{0.7}$As core - shell NWs in arrays with different pitch size $p$ (sample-1).  (scale bars correspond to 500 nm)}
	\label{fig:1_3}
\end{figure}

For the growth of the NW core and the shell we used a portable MBE chamber which is equipped with Ga and In effusion cells and an $As_4$ valve cracker cell as animated in figure \ref{fig:1_3}(a). 
It was observed in \cite{Hertenberger2010} that the NW volume depends on the number density of growing NWs explained by the differences in the local growth conditions. Considering the substrate area around an individual NW is acting as reservoir for NW growth, the decrement of $p$ associated with shadowing of the Ga-flux by neighbored NWs results in the growth of shorter and thinner NWs compared to the ones grown with larger $p$ (lower density)\cite{Plissard2010,Martensson2004,Sibirev2012,Oehler2018}.\\
An observed influence of the cells arrangement of the same pMBE chamber on the bending direction of the NWs was reported in \cite{Alhumaidi2021}. It was observed that NWs grown on patterned Si substrates with the given growth parameters bend toward the direction of the Ga flux (which is perpendicular to GaAs[11-2] lattice planes) in case of performing no substrate rotation during shell growth. The predetermination of the bending direction and the geometrical arrangement of the pMBE effusion cells with respect to the NW arrays are illustrated schematically in \ref{fig:1_3}(a).\\

The exploited shadowing effect of the growth material flux by neighboring NWs and the resulting bending profile along the NW is demonstrated in figure \ref{fig:1_3}(b). 
The NW arrays with high number density (i.e. $p = 100$ nm) that experience prominent flux shadowing display a lower straight (shadowed) segment and upper bent (exposed) segment of the NW (see figures \ref{fig:1_3}(b) and \ref{fig:1_3}(d)).
The length ratio of the exposed $(l_{exposed})$ and the shadowed $(l_{shadowed})$ NW segments can be controlled by changing the flux angle ($\Phi_{flux}$), NW length ($l$) and the distance ($p$) as animated in figure \ref{fig:1_3}(b). The length of the exposed NW segment $(l_{exposed})$ is given by

\begin{equation}
	l_{exposed} = \frac{p}{\tan(\Phi_{flux})} 
	\label{eqn:lexposed}
\end{equation} 

\noindent For the length of the bent segment $l_{bent}$, the diffusion of the shell materials on the NW surface from the exposed part toward the shadowed part of the NW must be considered. Therefore,  
\begin{equation}
	l_{bent} = l_{exposed}+l_{D} 
	\label{eqn:lbent}
\end{equation}       
\noindent where $l_{D} $ is the length of the NW segment at which the shell growth takes place only by the diffused material, i.e. no direct-flux deposition.          

\noindent The length $l_D$ indicates the effective diffusivity of group-III materials on the NW surface that accumulate and causes strain-induced NW bending.
Figure \ref{fig:1_3}(c) shows an exemplary 30° tilt view SEM images of a reference GaAs NWs in arrays with $p = 100, 200, 400 $ and $700$ nm, evidencing high NW yield.
In figure \ref{fig:1_3}(d), 30° title view SEM images are taken from four NW arrays of sample-1 with different $p$ showing significant variation of bending profiles along the NWs which implies the variation of strain distribution along these NWs. For the NWs arrays with $p =$ 100 nm and $p=$ 200 nm, the bending occurs only at the upper part of the NW that appears thicker in diameter compared to the straight lower part due to the shadowing effect on the later.  

\subsection*{XRD measurement}

In order to get deeper insight into the evolution of the bending and the strain along the NWs at different arrays, we performed an \textit{in-situ} XRD experiment. By scanning the sample across the X-ray beam while fulfilling Bragg's condition of the GaAs(111) reflection, the micro-fields of NWs were precisely located with respect to the diffractometer geometry. For each field with particular $p$ we monitored the evolution of NW bending as function of In$_{0.15}$Ga$_{0.85}$As shell growth time for sample-2. During the increment of the shell thickness the axial component of the lattice mismatch strain $\epsilon_{||}$ increases which in turn exerts a stronger bending force on the NW core.
Figure \ref{fig:2_3}(a) shows the 3D distribution of the GaAs(111) Bragg reflection in reciprocal space (reciprocal space map RSM) for the NW arrays with $p = 100, 200$ and $400$ nm recorded after different times of In$_{0.15}$Ga$_{0.85}$As shell growth. The recorded 3D RSMs shown in figure \ref{fig:2_3}(a) are represented by the reciprocal space vectors $Q_x$, $Q_y$ and $Q_z$ where $Q_z$ is set parallel with the GaAs[111] NW growth axis and sensitive to the axial strain $\epsilon_{||}$ while $Q_x$ and $Q_y$ are parallel with the Si(111) plane of the substrate and sensitive to the changes in the crystal orientation (i.e. tilting and bending of the NW). In addition to these vectors we introduce two new vectors, $Q$ and $Q_r$ where $Q$ set along the bending direction on $Q_xQ_y$ component of the RSM as demonstrated by blue arrow in figure \ref{fig:2_3}(a). The other vector $Q_r$ has the same origin as the other vectors and is tilted from $Q_z$ by bending angle of the NW crystal. A detailed depiction of the mentioned vectors as well as the method used for the strain calculation along the NWs can be found in \cite{Alhumaidi2021}. 
Additionally, in figure \ref{fig:2_3}(a) 2D cross-sections of the 3D Bragg reflections on $QQ_z$ component of the RSMs are highlighted. The lower images in figure \ref{fig:2_3}(a) are the GaAs(111) Bragg peak of straight GaAs NWs and the top images are that of NWs after 20 minutes of shell growth.

\section{Bending distribution along the NW}

\begin{figure}[!ht]
	\includegraphics[width=\textwidth]{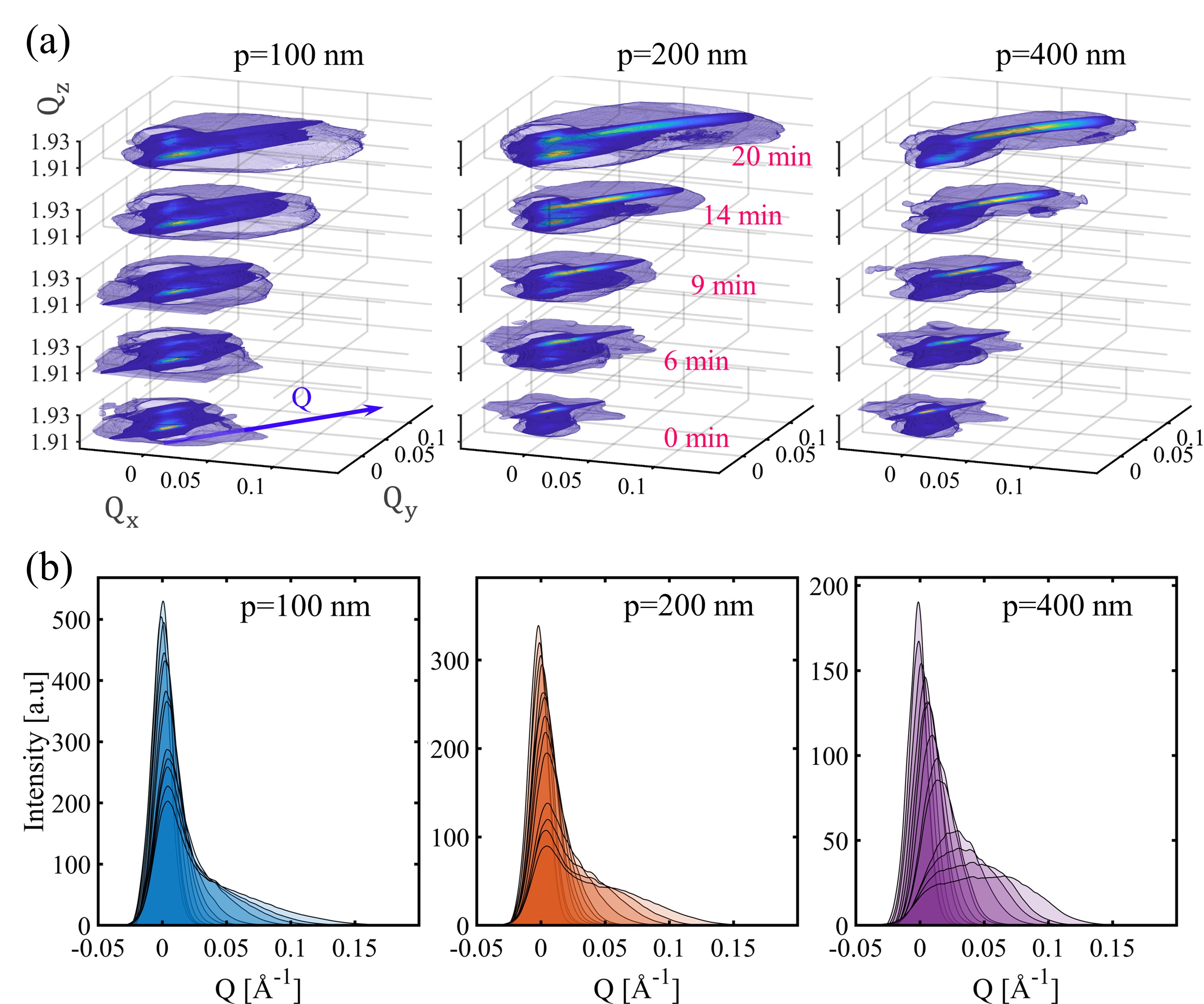}
	\caption{(a) 3D RSMs of the GaAs(111) Bragg's reflection after different shell growth rounds stacked vertically for NW arrays with $p=100, 200$ and $400$ nm. (b) Integrated line profile of the intensity distribution of the XRD signal along $Q$ in RSM.}
	\label{fig:2_3}
\end{figure}

The recorded signals in \ref{fig:2_3}(a) show that for the arrays with $p = 100$ nm and $p=200$ nm the most intense part of the Bragg’s reflection locate close to $Q$=0 $\AA ^{-1}$ with some broadening toward higher $Q$ values.
This finding implies that a large section of the lower part of the NWs remains straight and perpendicular to the substrate surface, while the broadened part reflects the bend section at the upper part of the NW. In contrast for the arrays with $p=400$ nm, the peaks exhibit rather homogeneous broadening toward higher ($Q$) values indicating that the whole NW bends.

\begin{figure}[!ht]
	\centering
	\includegraphics[width=\textwidth]{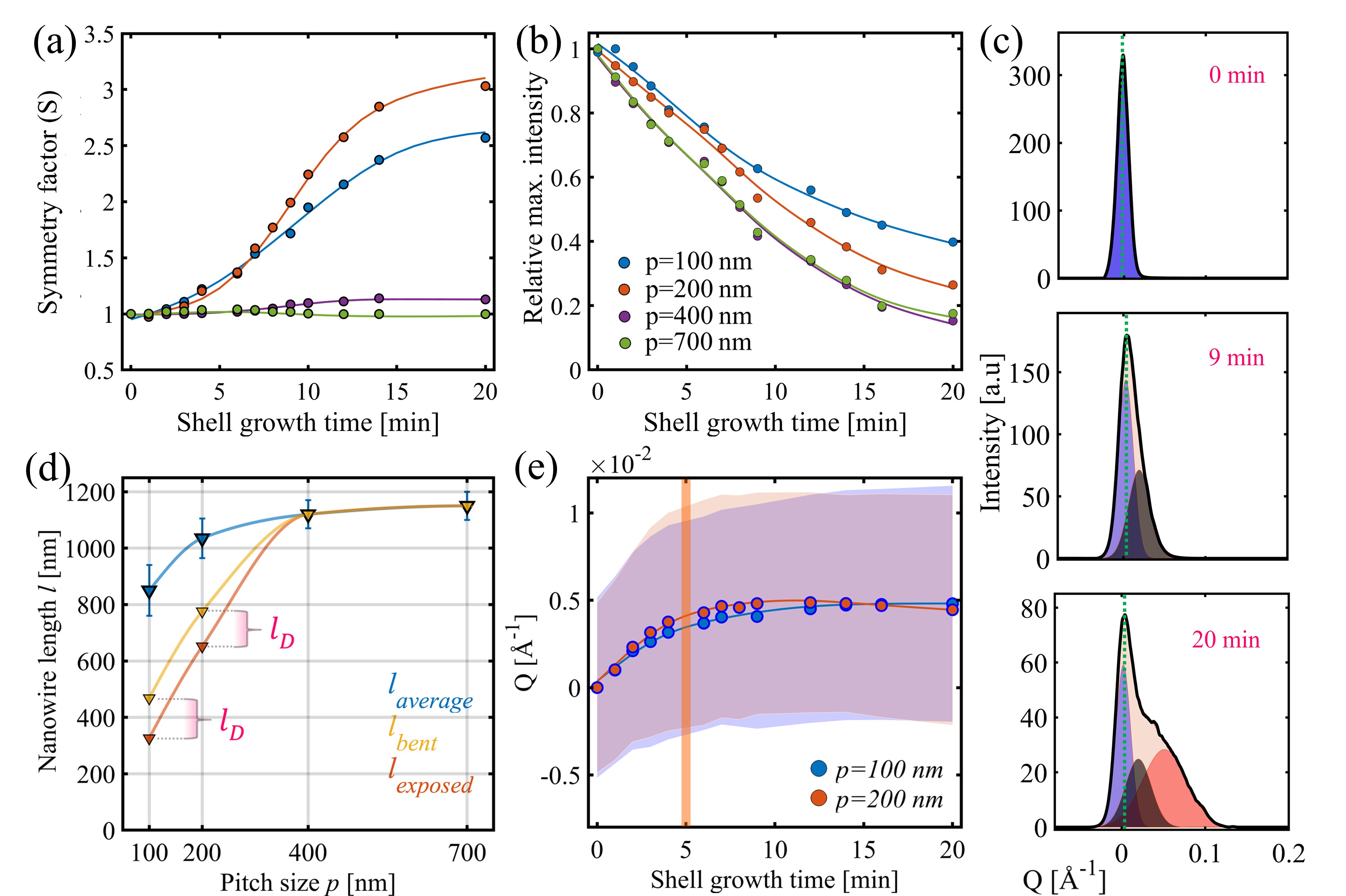}
	\caption{(a) The calculated symmetry factor $S$ of the line profiles of the GaAs(111) NW Bragg peaks for different NW arrays, integrated along $Q$ in RSM as a function of shell growth time. (b) The relative changes of maximum intensities of Bragg peaks of the NWs at different arrays as function of shell growth time. (c) Exemplary plots of the multi-Gaussian fitting model used to deconvolute the integrated line profiles of the Bragg peak of the NWs at the arrays with $p=$ 100 nm. (d) Estimated length of the NW segments that exhibited bending extracted from the XRD peak profile and compared to the length of the exposed segment indicating the length of the NW segment that is covered by the diffused materials ($l_{D}$). (e) The peak position of the XRD signal of the lower part of the NW at the arrays with $p=$ 100 nm and $p=200$ nm as function of shell growth time}.
	\label{fig:3_3}
\end{figure}

\noindent The integrated intensity profiles of Bragg peaks along $Q$ for the arrays with $p=100$, $200$ and $400$ nm are plotted in figure \ref{fig:2_3}(b). It can be seen that the intensity of the integrated line profiles depends on the number density of the NWs at the different arrays. The evolution of the Bragg peak intensity distribution along $Q$ shows that the peak maxima decreases and the peak shape is changing as function of shell growth time, forming a tail of the Bragg peak profile at the arrays with $p = 100$ nm and $p=200$ nm . 
The increasing tailing indicates the evolution of the NW curvature. \\
To quantify the signal profiles, we first introduce the symmetry factor $S$ (also known as tailing factor) of each Bragg peak along Q by 
\begin{equation}
	S=\frac{W_{0.05}}{2f_{0.05}}
\end{equation}
where $W_{0.05}$ is the peak width at the 5\% of the peak height and $f_{0.05}$ is the distance from the leading edge of the peak at 5\% of the peak height to the position of the peak maxima on $Q$. 
The evaluated $S$ values for each Bragg peak for four different NW arrays with different $p$ are plotted as function of shell growth time in figure \ref{fig:3_3}(a). It can be seen, $S$ of the Bragg peaks of the arrays with $p=$ 400 nm and $p=$ 700 nm during shell growth are close to $S\approx 1$. This high peak symmetry evidences that the entire NW experience curvature. In contrast, at the high density NW arrays with $p = 100$ nm and $p=200$ nm, $S$ becomes larger than unity caused by higher tailing of Bragg peaks toward higher $Q$ values which in turn results from the inhomogeneity of the NW curvature at these arrays. \\
\noindent Furthermore, as the NW curvature increases, the maximum intensity of Bragg peak $I_{max}^{p}(t)$ decreases due to the spreading of the diffracted signal distribution along $Q$. Figure \ref{fig:3_3}(b) shows the relative changes of the Bragg peak maxima at different shell growth times.

The maxima of the Bragg peak intensity of the bare GaAs NW (which locates around $Q=0\AA^{-1}$) $I_{max}^{p}(0)$ drops to $\approx 20\%$ after 20 minutes of shell growth at the arrays with $p=400$ nm and $p=700$ nm. Whilst, the maxima of Bragg peak intensity profile of the NW arrays with $p=100$ nm decreases to $\approx 45\%$ and $\approx 25\%$ for the arrays with $p=200$ nm, respectively. These values imply that percentage of the diffracted signals from the NWs accumulate at the same position in the RSM for these two arrays. Therefore, approximately 45\% and 25\% of the NW volume remained vertical to the substrate surface for the NW arrays with $p=100$ nm and $p=200$ nm, respectively. 
To innervate this approach, the Bragg peaks of the bent NW at these arrays were deconvoluted by multiple-Gaussians as shown in figure \ref{fig:3_3}(c). 
By integrating the area of each Gaussian, this model gives the same percentages of the XRD signal that remained close to $Q=0\AA^{-1}$ in RSM for the tow mentioned arrays as shown before (shaded with blue in figure \ref{fig:3_3}(c)).

\noindent Accordingly, the average volume of the NW part that exhibit bending forms about 55\% and 75\% of the total volume of the NWs at the arrays with $p=100$ nm and $p= 200$ nm, respectively.
\begin{table}[!h]
	\begin{center}
		\caption{Measured average length $l_{average}$ of the NWs at different arrays and the calculated length of the shell-material exposed segment $l_{exposed}$.}
		\label{tab:table1}
		\begin{tabularx}{9cm}{ccc}
			\toprule
			Pitch size ($p$) [nm] &  $l_{average}$ [nm]   &   $l_{exposed}$ [nm]   \\ 
			
			\hline
			\midrule 
			100  &	850  ± 90&  320 \\
			200  &	1035 ± 70&   652\\
			400  &	1120 ± 50&	1120\\
			700  &	1150 ± 50&	1150\\
			1000  &	1150 ± 50&	1150\\
			
			\bottomrule
			
		\end{tabularx}
	\end{center}
\end{table}

\noindent The results are indicating that the NWs are not homogeneously covered by the shell material which is caused by the shadowing of the material flux by neighbored NWs. The length of the exposed segment of the NW can be calculated from equation \ref{eqn:lexposed} as listed in table \ref{tab:table1} by considering the distance between the NWs which is $\sqrt{3}\:p$ in our case.

\noindent By considering the average length of the NWs $l_{average}^p$ for the mentioned arrays, the length of the bent part $l_{bent}^p$ can be calculated by 
$l_{bent}^{100}=0.55\times l_{average}^{100}=468$ nm and $l_{bent}^{200}=0.75\times l_{average}^{200 }=776$ nm.  
The values of $l_{average}$, $l_{bent}$ and $l_{exposed}$ are plotted in figure \ref{fig:3_3}(d). From these values, one can estimate the length of the NW segment $l_D^p$ that is covered by the diffused shell materials by $l_D^{100}=l_{bent}^{100}-l_{exposed}^{100}\approx 142$ nm and $l_D^{200}=l_{bent}^{200}-l_{exposed}^{200}\approx 124$ therefore $l_{D}^{average}\approx 135$ nm. Therefore equation \ref{eqn:lbent} can be modified by

\begin{equation}
	l_{bent} = \frac{p}{\tan(\Phi_{flux})} +135
\end{equation}   

\noindent The length $l_{D}$ however depends on the diffusivity of the shell materials which in turn depends on the NW surface properties as well as the growth temperature and the V-III ratio (i.e. As pressure in our case). Therefore, the length $l_{D}$ is valid for the given parameters of the shell growth of studied sample and may be changed by tuning these parameters.  \\

\noindent Figure \ref{fig:3_3}(e) shows the position of the XRD peak maxima taken from $p=$ 100 nm and $p=$ 200 nm arrays as function of shell growth time. It can be seen that the peak position exhibited minor changes during the first 5 minutes indicating small NW bending of$\approx$ 0.025$^\circ$, under the consideration of the angular resolution of our measurement being limited to 0.01$^\circ$ by the XRD setup. However, the minor changes of the peaks position during the first 5 minutes of the shell growth indicate a minor development of the strain and the curvature in the entire NW at the beginning. This might be explained by a high diffusivity of the shell material at the early stages of shell growth and may decrease as the shell thickness increase and the strain on the NW surface increases. 

\section{Strain distribution along the NW}

\begin{figure}[!ht]
	\includegraphics[width=\textwidth]{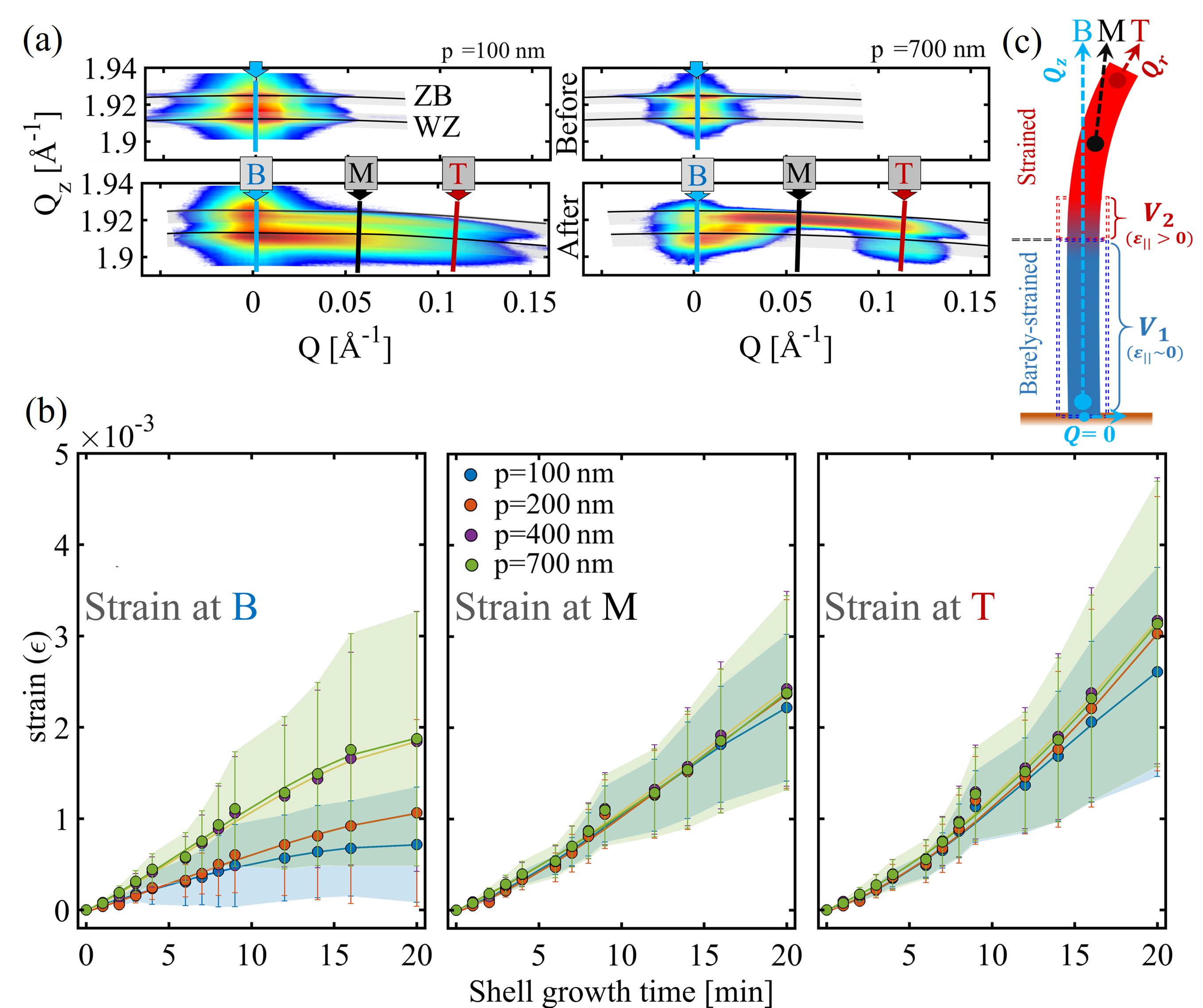}
	\centering
	\caption{(a) 2D section of XRD peak of GaAs(111) in RSM on $QQ_z$ of the arrays with $p=$ 100 nm and $p=$ 700 nm of the bare NWs and after 20 minutes of shell growth where the B,M and T labeled arrows indicate the positions of the line profiles along $Q_r$ used for strain calculation. (b) The extracted average strain values indicated by symbols and the strain variation indicated by the shaded area at different positions along the NW plotted as function of shell growth time. (c) Illustration of the strain distribution along the NW and the corresponding positions of B,M and T as well as the volume of the lower NW part and the volume of the segment that may be included in the overlapping of the XRD signal when integrating a line profile at B.}
	\label{fig:5_3}
\end{figure}

Benefiting from the spatial distribution of the XRD signal in reciprocal space of the bent NW, we are able to measure the average strain at different parts of the NW during shell growth. By profiling the XRD signal along $Q_{z}$ and $Q_r$ at different positions on $Q$ determined from the multiple-Gaussian fitting shown in figure \ref{fig:3_3}(c) for the arrays with $p=$ 100 nm and $p=$ 200 nm. The line profiles along $Q_z$ are taken at $Q\approx 0\AA^{-1}$ of GaAs(111) NWs at the different arrays (indicated by blue lines in figure \ref{fig:5_3}(a)). At this his position in the RSM the diffracted signal from the lowest part of the NW (i.e. NW bottom, we denote it in the text by B) takes place. To evaluate the strain at the middle segment (denoted by M) of the bent part of the NW we integrate line profiles along $Q_r$ for each recorded RSM. The line profile along $Q_r$ in this case is taken at the peak shoulder along $Q$ in case of the arrays with $p=$ 100 nm and $p=$ 200 nm, and at the peak center for the arrays with $p=$ 400 nm and $p=$ 700 nm as indicated by black lines in figure \ref{fig:5_3}(a). 

By considering the distribution of the zinc-blende (ZB) and wurtzite (WZ) structures along the grown GaAs NWs, 
the strain at the top part of the NW (denoted by T) could be evaluated. For ZB phase dominated NWs, WZ appears at the top and the bottom of the NW due to the changes of the growth conditions during the axial growth of the NWs as reported in \cite{Alhumaidi2021}. Therefore, we integrate a line profile along $Q_r$ at the position of the displaced WZ peak on $Q$ for the arrays with $p=$ 400 nm and $p=$ 700 nm while we consider the peak tail on $Q$ for the arrays with $p=$ 100 nm and $p=$ 200 nm as indicated by red arrows in figure \ref{fig:5_3}(a).  

\noindent The obtained average strain $\epsilon_{||}$ is measured from the peak position on $Q_r$ or $Q_z$ and the strain variation $	\Delta \epsilon_{||}$ is obtained from the peak broadening on the same vectors as explained in details \cite{Alhumaidi2021} by the following formulas

\begin{equation}
	\epsilon_{||}=\frac{Q^t_r-Q^0_r}{Q^0_r}
	\label{strain}
\end{equation} 

\begin{equation}
	\Delta \epsilon_{||}=\frac{\Delta (Q^t_r)-\Delta (Q^0_r)}{\Delta (Q^0_r)}
	\label{strainVariation}
\end{equation}
where $\Delta (Q^t_r)=(Q^t_r-\sigma^t_r)$ takes into account the standard deviation $\sigma^t_r$ of the signal along $Q_r$, $Q^t_r$ is the peak center on $Q_r$ at time $t$ and $Q^t_0$ is the peak center before shell growth. 
From equations \ref{strain} and \ref{strainVariation} $\epsilon_{||}$ and $	\Delta \epsilon_{||}$ at the mentioned positions B, M and T of the NWs are plotted for the different arrays as a function of the shell growth time in figure \ref{fig:5_3}(b). As it can be seen, the strain value changes as the shell growth time increases in different manners for the different parts of the NWs at the different arrays. The different manners of strain evolution can be sorted as following:
\begin{itemize}
	\item \textbf{At the lower part B of the NW}, the average strain $\epsilon_{||}^{B,p}$ in the NWs of the arrays with $p=$ 100 nm increases and saturates at $\epsilon_{||}^{B,100}=$0.0007 after 16 minutes of shell growth while $\epsilon_{||}^{B,200}$ increase in a nonlinear fashion and reaches a value of $\epsilon_{||}^{B,200}=$0.001 after 20 minutes of shell growth as shown in left panel of figure \ref{fig:5_3}(b). This approach indicates the low strain magnitude at the lower parts of the NWs where the shadowing effect takes place. However, the minor strain that built-up in the shadowed part of the NWs at these arrays is in logical consistency with the observation of the slight bending as shown in figure \ref{fig:3_3}(e). In this case, the entire NW at first few minuted of shell growth is slightly strained due to the high diffusivity of the shell material on the NW surface where this diffusivity decreases as the strain increases. The same approach explains the saturated strain variation as shaded with blue in the left panel of figure \ref{fig:5_3}(b).
	Additionally, an overlapping of the signal in the RSM of the bent (strained) part of the NW with the lower (barely-strained) part must be considered as a contributor to the strain curve plotted at these arrays. The volume of the segments that contribute to the overlapping is demonstrated by the red rectangle in figure \ref{fig:5_3}(c) and denoted by $V_2$, where $V_1$ is the volume of the straight part of the NW. As the NW curvature increases, the XRD signal from the entire NW spreads more along $Q$ in the RSM and the overlapping of the signals from the different NW segments decreases; therefore, $\frac{V_2}{V_1}$ decreases. The ratio $\frac{V_2}{V_1}$ explains the relatively higher strain values at B of the arrays with $p=$ 200 nm comparing to the ones of arrays with $p$=100 nm, where $V_1$ forms 25\% of the NW volume at the array with $p=$ 200 nm comparing to 45\% for the ones with $p=$ 100 nm as shown above.

	\noindent For the NW arrays with $p=$ 400 nm and $p=$ 700 nm, the average strain increases in the same manner for both NW arrays and reaches a value of $\epsilon_{||}^{B,400}=\epsilon_{||}^{B,700}=$ 0.0019 after 20 minutes of shell growth. The strain variation of these arrays increase as the shell growth proceeds as indicated by the green shade in the left panel of figure \ref{fig:5_3}(b). 
	However, it was reported in literatures that the strain magnitude at the NW base near the wire-substrate interface is relatively low comparing to the other parts of the NW \cite{Gronqvist2010,Davtyan2020}. This feature at the NW base explains the curve shape of the strain function of the arrays with $p=400$ nm and $p=$ 700 nm plotted in left panel of figure \ref{fig:5_3}(b). The same feature explains the higher strain variation at B for the same arrays comparing to the other parts M and T that are shaded in green in figure \ref{fig:5_3}(b) by the same concept of the overlapping signals in RSMs.  
	
	\item \textbf{At the position M on the NW}, the average strain magnitude of all NW arrays increases to $\epsilon_{||}^{M,200}=\epsilon_{||}^{M,400}=\epsilon_{||}^{M,700}=$ 0.0024 while $\epsilon_{||}^{M,100}=$ 0.0022 as the shell growth proceed as well as the strain variation as it can be seen in middle panel of figure \ref{fig:5_3}(b). This implies a progressive strain evolution and an increasing asymmetry degree of the shell growth around the NW at the parts where the flux shadowing takes no place.
	
	\item \textbf{At the position T at the NW top}, the strain increases as a quadratic function to the shell growth time and reaches a higher value comparing to the strain at C where $\epsilon_{||}^{M,100}=$0.0026, $\epsilon_{||}^{M,200}=$ 0.003 and $\epsilon_{||}^{M,400}=\epsilon_{||}^{M,700}=$ 0.0031 as it can be seen in right panel of figure \ref{fig:5_3}(b). This increment of the strain may be explained by the changes of the local deposition geometry of the shell growth material along the NW. As the NW curves, the angle of the incident flux changes along the NW as demonstrated in \cite{Spencer2021} which leads to an inhomogeneous shell thickness along the exposed segment of the NW. 
	However, in our study, we relate to the early stages of NW bending and the maximum bending angle doesn't exceed 5$^\circ$ at the NW tip, therefore, the variation of the strain magnitude at B, M and T would increase as the NW bending increases.

\end{itemize}

\subsubsection{Conclusion}
In this work we employed the knowledge about the preferable bending direction of the NWs with respect the azimuthal arrangement of the MBE cells as well as the pitch size to control the shell material distribution along the NWs for different arrays. The strain distribution and the subsequent bending profile of the NW were different at different NW arrays with different densities. At the arrays with high NW densities, the shell materials are deposited asymmetrically on the NW part that is exposed to the direct flux. The diffusion of the shell material toward the shadowed part of the NW was considered and the whole mechanism results in a strained and bent upper part of the NW while the lower part remains rather straight and barely strained. \\
At theses NW arrays we could estimate the length of the NW segment that is strained by the diffused shell materials by means of \textit{in-situ} X-ray diffraction. Additionally, we observed that the shadowed part of the NW exhibits a low strain magnitude and minor bending during the early stages of shell growth, indicating high diffusivity of the shell materials at the beginning.  
At the NW arrays with low densities where the shadowing effect takes no place, the shell materials cover the whole length of the NW. At these arrays the entire NW exhibit bending and the XRD measurement revealed that the strain evolve in the same manner at different parts of the NW. 
These results provide recipes for controlling the NW geometry with novel designs which might be used for NWs interconnect as well as tuning the strain distribution along the NW.  

\subsubsection{Acknowledgements}
The authors thank B. Krause, A. Weisshardt, J. Kalt and S. Stankov for their support at KIT, as well as the INT for access to the SEM. We acknowledge DESY (Hamburg, Germany), a member of the Helmholtz Association HGF, for the provision of experimental facilities. Parts of this research were carried out at PETRA III and we would like to thank A. Khadiev at In situ X-ray diffraction and imaging beamline P23 and M. Lippmann for help in the Clean Room. Other parts of this research were carried out at the DESY Nanolab and we would like to thank T. F. Keller, A.  and S. Kulkani for using the SEM. We are grateful to O. Kr\"{u}ger and M. Matalla (Ferdinand-Braun-Institute, Berlin) for electron beam lithography, and to S. Meister and S. Rauwerdink for substrate preparation. This work was funded by BMBF project 05K16PSA and partially supported by Deutsche Forschungsgemeinschaft under grant Ge2224/2.

\bibliography{ShadowingEffectNWs2}

\end{document}